# Phases vs. Levels using Decision Trees for Intrusion Detection Systems

Heba Ezzat Ibrahim, Sherif M. Badr and Mohamed A. Shaheen
College of Computing and Information Technology
Arab Academy for Science, Technology and Maritime Transport
Cairo, Egypt
Heba_ezzat_86@yahoo.com

*Abstract*— **Security of computers and the networks that connect them is increasingly becoming of great significance. Intrusion detection system is one of the security defense tools for computer networks. This paper compares two different model Approaches for representing intrusion detection system by using decision tree techniques. These approaches are Phase-model approach and Level-model approach. Each model is implemented by using two techniques, New Attacks and Data partitioning techniques. The experimental results showed that Phase approach has higher classification rate in both New Attacks and Data Partitioning techniques than Level approach.**

## I. INTRODUCTION

The Internet and online procedures is an essential tool of our daily life. They have been used as a main component of business operation [1]. Therefore, network security needs to be carefully concerned to provide secure information channels [2].

It is difficult to prevent attacks only by passive security policies, firewall, or other mechanisms. Intrusion Detection Systems (IDS) have become a critical technology to help protect these systems as an active way. An IDS can collect system and network activity data, and analyze those gathered information to determine whether there is an attack [3].

Network Intrusion detection (NIDS) and prevention systems (NIPS) serve a critical role in detecting and dropping malicious or unwanted network traffic [5]. Intrusion detection and prevention systems (IDPS) are primarily focused on identifying possible incidents, logging information about them, attempting to stop them, and reporting them to security administrators. In addition, organizations use IDPSs for other purposes, such as identifying problems with security policies, documenting existing threats, and deterring individuals from violating security policies. IDPSs have become a necessary addition to the security infrastructure of nearly every organization [6].

Intrusion detection started in around 1980s after the influential paper from Anderson [4]. Intrusion detection systems are classified as network based, host based, or application based depending on their mode of deployment and data used for analysis [7]. Additionally, intrusion detection systems can also be classified as signature based or anomaly based depending upon the attack detection method. The signature-based systems are trained by extracting specific patterns (or signatures) from previously known attacks while the anomaly-based systems learn from the normal data collected when there is no anomalous activity [7].

Another approach for detecting intrusions is to consider both the normal and the known anomalous patterns for training a system and then performing classification on the test data. Such a system incorporates the advantages of both the signature-based and the anomaly-based systems and is known as the Hybrid System. Hybrid systems can be very efficient, subject to the classification method used, and can also be used to label unseen or new instances as they assign one of the known classes to every test instance. This is possible because during training the system learns features from all the classes. The only concern with the hybrid method is the availability of labeled data. However, data requirement is also a concern for the signature-based and the anomaly-based systems as they require completely anomalous and attack free data, respectively, which are not easy to ensure [8].

## II. PREVIOUS WORK

The purpose of IDS is to help computer systems with how to discover attacks, and that IDS is collecting information from several different sources within the computer systems and networks and compares this information with preexisting patterns of discrimination as to whether there are attacks or weaknesses [10].

Decision Trees (DT) have also been used for intrusion detection [11]. Decision Tree is very powerful and popular machine learning algorithm for decision-making and classification problems. It has been used in many real life applications like medical diagnosis, radar signal classification, weather prediction, credit approval, and fraud detection etc [12]. The decision tree is a simple if then else rules but it is a very powerful classifier and proved to have a high detection rate. They are used to classify data with common attributes. Each decision tree represents a rule which categorizes data according to these attributes. A decision tree has three main components: nodes, leaves, and edges. Each decision tree represents a rule set, which categorizes data according to the attributes of dataset. The DT building algorithms may initially build the tree and then prune it for more effective classification. [13].

### A. C5.0 Decision Trees

See5.0 (C5.0) is one of the most popular inductive learning tools originally proposed by J.R.Quinlan as C4.5 algorithm (Quinlan, 1993) [13].

C5.0 can deal with missing attributes by giving the missing attribute the value that is most common for other instances at the same node. Or, the algorithm could make probabilistic calculations based on other instances to assign the value [14].

### B. Classification and Regression Trees (CRT or CART)

CART is a recursive partitioning method to be used both for regression and classification. The key elements of CART analysis are a set of rules for splitting each node in a tree; deciding when tree is complete and assigning a class outcome to each terminal node. CART is constructed by splitting subsets of the data set using all predictor variables to create two child nodes repeatedly, beginning with the entire data set [15].

### C. Chi-squared Automatic Interaction Detector (CHAID)

CHAID (Chisquare-Automatic-Interaction-Detection) was originally designed to handle nominal attributes only.
CHAID method is based on the chi-square test of association. A CHAID tree is a decision tree that is constructed by repeatedly splitting subsets of the space into two or more child nodes, beginning with the entire data set [16].
CHAID handles missing values by treating them all as a single valid category. CHAD does not perform pruning.

### D. Quick, Unbiased, Efficient Statistical Tree (QUEST)

QUEST is a binary-split decision tree algorithm for classification and machine learning. QUEST can be used with univariate or linear combination splits. A unique feature is that its attribute selection method has negligible bias. If all the attributes are uninformative with respect to the class attribute, then each has approximately the same change of being selected to split a node [17].

We compare between the phase model in [9], and the Level model in [6].The authors in [9] design a system which consists of three detection levels. The network data are introduced to the module of the first level which aims to differentiate between normal and attack. If the input record was identified as an attack then the administrator would be alarmed that the coming record is suspicious and then this suspicious record would be introduced to the second level which specifies the class of this attack (DOS, probe, R2L or U2R). The third detection level consists of four modules one module for each class type to identify attacks of this class. Finally the administrator would be alarmed of the expected attack type.

In [6], the authors classify network intruders into a set of different levels. The first level is called the Boolean detection level, where the system classifies the network users to either normal or intruder. The second level is called the coarse detection level, where it can identify four categories of intruders. The third level is called the fine detection level, where the intruder types can be fine tuned into 23 intruder types.

### III. SYSTEM ARCHITECTURE

*The system components :*

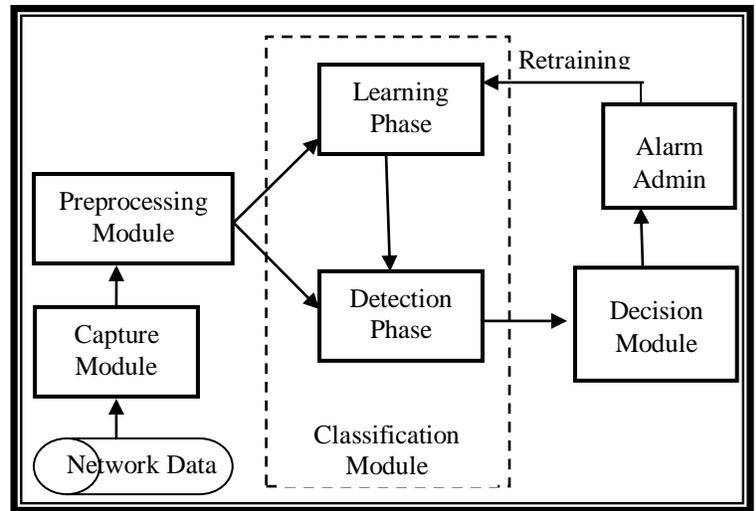

Figure 1. System components

*Figure 1. shows the main modules of IDS as follows:*

### A. The Capture Module

Raw data of the network are captured and stored using the network adapter. It utilizes the capabilities of the TCP dump capture utility for Windows to gather historical network packets.

### B. The Preprocessing Module

The data must be of uniform representation to be processed by the classification module. The preprocessing module is responsible for reading, processing, and filtering the audit data to be used by the classification module. The preprocessing module handles Numerical Representation, Normalization and Features selection of raw input data. The preprocessing module consists of three phases: [18]

*1) Numerical Representation:* Converts non-numeric features into a standardized numeric representation. This process involved the creation of relational tables for each of the data type and assigning a number to each unique type of element. (e.g. protocol_type feature is encoded according to IP protocol field: TCP=0, UDP=1, ICMP=2). This is achieved by creating a transformation table containing each text/string feature and its corresponding numeric value.

*2) Normalization:* The ranges of the features were different and this made them incomparable. Some of the features had binary values where some others had a continuous numerical range (such as duration of connection). As a result, inputs to the classification module should be scaled to fall between zero and one [0, 1] range for each feature.[9]

*3) Dimension reduction:* reduce the dimensionality of input features of the classification module. Reducing the input dimensionality will reduce the complexity of the classification module, and hence the training time.

### C. The classification Module

The classification module has two phases of operation. The learning and the detection phase.

#### 1) The Learning Phase

In the learning phase, the classifier uses the preprocessed captured network user profiles as input training patterns. This phase continues until a satisfactory correct classification rate is obtained.

#### 2) The Detection Phase

Once the classifier is learned, its capability of generalization to correctly identify the different types of users should be utilized to detect intruder. This detection process can be viewed as a classification of input patterns to either normal or attack.

### D. The Decision Module

The basic responsibility of the decision module is to transmit an alert to the system administrator informing him of coming attack. This gives the system administrator the ability to monitor the progress of the detection module.

To evaluate our system we used two major indices of performance. We calculate the detection rate and the false alarm rate according to the following assumptions [19]:

- False Positive (FP): the total number of normal records that are classified as anomalous
- False Negative (FN): the total number of anomalous records that are classified as normal
- Total Normal (TN): the total number of normal records
- Total Attack (TA): the total number of attack records
- Detection Rate = [(TA-FN) / TA]*100
- False Alarm Rate = [FP/TN]*100
- Correct Classification Rate = Number of Records Correctly Classified / Total Number of records in the used dataset

There are four major categories of networking attacks. Every attack on a network can be placed into one of these groupings [20].

*1) Denial of Service Attack (DoS):* is an attack in which the attacker makes some computing or memory resource too busy or too full to handle legitimate requests, or denies\ legitimate users access to a machine.

*2) User to Root Attack (U2R):* is a class of exploit in which the attacker starts out with access to a normal user account on the system (perhaps gained by sniffing passwords, a dictionary attack, or social engineering) and is able to exploit some vulnerability to gain root access to the system.

*3) Remote to Local Attack (R2L):* occurs when an attacker who has the ability to send packets to a machine over a network but who does not have an account on that machine exploits some vulnerability to gain local access as a user of that machine.

*4) Probing Attack:* is an attempt to gather information about a network of computers for the apparent purpose of circumventing its security controls

*Two different model Approaches are built for intrusion detection system (Phase-model approach and Level-model approach) that are defined as follows:*

#### 1) Phase-Model Approach

Phase model consists of three detection phases. The data is input in the first phase which identifies if this record is a normal record or attack. If the record is identified as an attack then the module inputs this record to the second phase which identifies the class of the coming attack. The second Phase module passes each attack record according to its class type to phase 3 modules. Phase 3 consists of 4 modules one for each class type (DOS, Probe, R2L, U2R). Each module is responsible for identifying the attack type of coming record.

Each Phase was examined with different Decision Tree techniques. The Three Phases are dependent on each other. In other word Phase 2 cannot begin until Phase 1 is finished. This approach has the advantage to flag for suspicious record even if attack type of this record wasn't identified correctly.

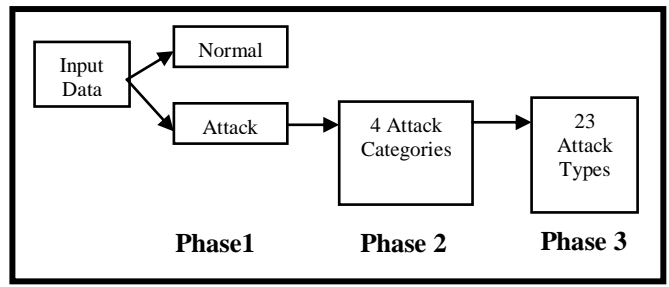

Figure 2. Phase Model Architecture

*2) Level-Model Approach*

Level model consists of 3 independent detection levels. The First Level is to detect normal and Attack profiles. The Second Level is to detect normal records and classify the attacks into four categories independently on the results of the first level. The third Level is to classify each attack type and normal records. Level model approach is to implement each level independent on the other level.

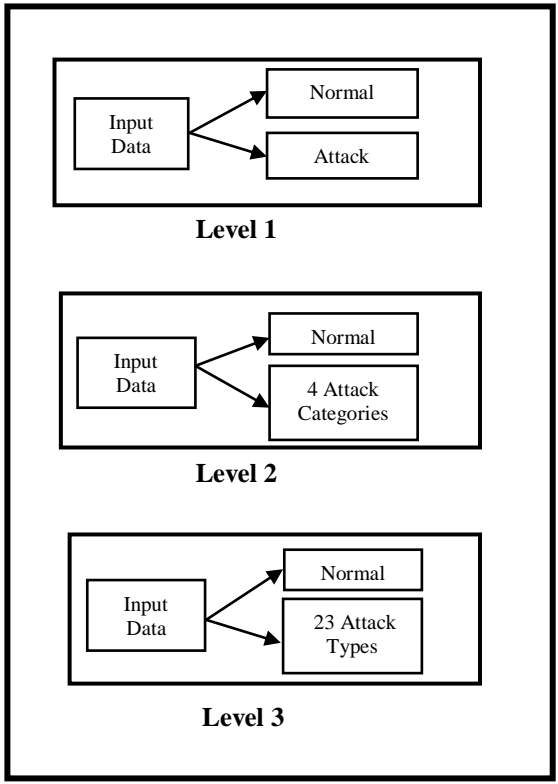

**Level 1**

**Level 2**

**Level 3**

Figure 3. Level Model Architecture

## IV. EXPERIMENTS AND RESULTS

### A. Data Description

KDDCUP'99 is the mostly widely used data set for the evaluation of these systems. The KDD Cup 1999 uses a version of the data on which the 1998 DARPA Intrusion Detection Evaluation Program was performed. They set up environment to acquire raw TCP/IP dump data for a local area network (LAN) simulating a typical U.S. Air Force LAN.

There are some inherent problems in the KDDCUP'99 data set [21], which is widely used as one of the few publicly available data sets for network-based anomaly detection systems

The data in the experiment is acquired from the NSLKDD dataset which consists of selected records of the complete KDD data set and does not suffer from mentioned shortcomings by removing all the repeated records in the entire KDD train and test set, and kept only one copy of each record [20]. Although, the proposed data set still suffers from some of the problems and may not be a perfect representative of existing real networks, because of the lack of public data sets for network-based IDSs, but still it can be applied as an effective benchmark data set to help researchers compare different intrusion detection methods. The NSL-KDD dataset is available at [22].

We used attacks from the four classes to check the ability of the intrusion detection system to identify attacks from different categories.

The two approaches are examined by two techniques:

*1) Test with New Attack:* The sample dataset contains 83644 record for training (40000 normal and 43644 for attacks) and 19784 for testing (9647 normal, 6935 for known attacks and 3202 for unknown attacks).

*2) Test by Data Partitioning:* The sample dataset contain 103427 records is partitioned by 10% (10156 records) for training and 90% (93271 records) for testing.

### B. Phase-Module Approach Results

*1) Test with New Attack:*

Results of Phases model tested with new attacks showed that C5 has a significant detection rate for known and unknown attacks in all phases.

TABLE I. Classification Rate of Phases with New Attacks

| Classifier | Correct Classification Rate | | |
|---|---|---|---|
| | Phase 1 | Phase 2 | Phase 3 |
| **C5** | 100 % | **85.34 %** | **99.32%** |
| **CRT** | 100 % | 83.62 % | 97.55% |
| **Chaid** | 100 % | 85% | 98.73% |
| **Quest** | 100 % | 73.11 % | 93.48% |

*2) Test by Data Patitioning:*

Results of data partitioning showed that C5 then CRT & CHAID produced best correct classification rate in second phase which is responsible for classifying coming attack to one of the four classes (DOS, Probe, R2L & U2R). In third phase, C5 showed it has the best classification rate as shown in table II.

TABLE II Classification Rate of Phase with Data Partitioning

| Classifier | Correct Classification Rate | | |
|---|---|---|---|
| | Phase 1 | Phase 2 | Phase 3 |
| C5 | 100 % | **99.98 %** | **99.49%** |
| CRT | 100 % | 99.97 % | 97.02 % |
| Chaid | 100 % | 99.79 | 97.38 % |
| Quest | 100 % | 93.74 % | 93.25 % |

Phase-Model approach has Detection Rate equal to 100 % in both New Attack and Data Partitioning techniques as all attacks in phase 1 are detected correctly.

*C. Level-Module Approach Results*

*1) Test with New Attack:*
Testing results showed that C5 produced best correct classification rate for third level and Quest for second level as shown in table III.

TABLE III Classification Rate of Levels with New Attacks

| Classifier | Correct Classification Rate | | |
|---|---|---|---|
| | Level 1 | Level 2 | Level 3 |
| C5 | 100 % | 83.82 % | **83.61 %** |
| CRT | 100 % | 91.72 % | 82.87 % |
| Chaid | 100 % | 83.64 % | 74.09 % |
| Quest | 100 % | **91.85 %** | 77.42 % |

TABLE IV Detection Rate of Levels with New Attacks

| Classifier | Detection Rate | | |
|---|---|---|---|
| | Level 1 | Level 2 | Level 3 |
| C5 | 100 % | 68.42 % | 100 % |
| CRT | 100 % | 100 % | 100 % |
| Chaid | 100 % | 68.41 % | 93.42 % |
| Quest | 100 % | 100 % | 100 % |

*2) Test by Data Patitioning:*
Results of data partitioning showed that second level are easy to be correctly classified by many decision trees classifiers either C5, CRT or CHAID. In third phase, C5 showed it has the best classification rate as shown in table V.

TABLE V Classification Rate of Levels with Data Partitioning

| Classifier | Correct Classification Rate | | |
|---|---|---|---|
| | Level 1 | Level 2 | Level 3 |
| C5 | 100 % | **99.96 %** | **99.73 %** |
| CRT | 100 % | 99.89 % | 90.22 % |
| Chaid | 100 % | 99.88 % | 87.92 % |
| Quest | 100 % | 97.17 % | 88.28 % |

TABLE VI Detection Rate of Levels with Data Partitioning

| Classifier | Detection Rate | | |
|---|---|---|---|
| | Level 1 | Level 2 | Level 3 |
| C5 | 100 % | 99.92 % | 100 % |
| CRT | 100 % | 100 % | 100% |
| Chaid | 100 % | 99.92 % | 96.52 % |
| Quest | 100 % | 100 % | 100 % |

## V. DISCUSSION

We defined two different Approaches. The first approach is the phase model approach which consists of three sequential detection levels. Phase 1 is able to detect Normal and Attack behavior. Phase 2 is to classify the attacks detected from phase 1 into 4 Attack categories (DOS, Probe, R2L, U2R). Phase 3 is to classify each attack type in each category.

The second approach is the level model approach which consists of 3 separated detection level. Level1 is to detect normal and Attack profiles. Level2 is to detect normal records and classify the attacks into four categories. Level3 is to classify each attack type and normal records.

TABLE VII Comparison between Phase and Level approaches

| | Phase Approach | Level Approach |
|---|---|---|
| Training Time | less training time | High training time |
| Detection Rate | Higher detection Rate for New Attacks | Lower detection rate for New Attacks |
| False Alarm Rate (FAR) | Lower FAR as Attacks are detected in the first phase | Higher FAR as Attacks Types and Categories a are detected in parallel with the normal records |
| Errors Propagation | May propagate errors | Does not propagate errors |
| Classification Rate | Higher Classification Rate in New Attacks and Data Partitioning Techniques | Lower classification Rate in New Attacks technique. |

As shown in table VII, Phase model take less training time and even decrease in each phase where we use the whole dataset for training phase 1 then in phase 2 we use only the attacks for training excluding the normal records. While in Level model, it takes high training time as the whole data is entered in the training of each level.

Phase model has higher detection Rate for New Attacks which never been seen before but lower detection rate for New Attacks in level model.

Attacks are detected in the first phase then are sent for further classification to the next phase without Normal records but in Level model, Attacks Types and Categories are detected in parallel with the normal records which may increase the false alarm rate.

Phase model May propagate errors as each phase is dependent on the previous one. But level model does not propagate errors as each level is separated and has independent results.

Phase model has Higher Classification Rate in New Attacks and Data Partitioning Techniques than Level model which has Lower classification Rate in New Attacks technique.

## VI.  CONCLUSION AND FUTURE WORK

In this paper we compared the results of 2 different approaches of intrusion detection system (Phase and Level Approach). Phase Approach consists of three detection phases. The data is input in the first phase which identifies if this record is a normal record or attack. If the record is identified as an attack then the module inputs this record to the second phase which identifies the class of the coming attack. The second phase module passes each attack record according to its class type to phase 3 modules. Phase 3 consists of 4 modules one for each class type (DOS, Probe, R2L, U2R). Each module is responsible for identifying the attack type of coming record. While the Level approach consists of 3 independent detection levels. The First Level is to detect normal and Attack profiles. The Second Level is to detect normal records and classify the attacks into four categories independently on the results of the first level. The third Level is to classify each attack type and normal records.

We examined each model approach using different decision trees modules (C5, CRT, QUEST and CHAID). Each module is implemented by applying 2 techniques (New Attacks and Data Partitioning Techniques) .First, New Attacks Technique is is to add new attacks in testing. Second, Data Partitioning Technique is to divide the dataset into 10 %for training and 90% for testing.
New Attacks technique is more realistic than Data Partitioning technique as in real life we are exposed to new attacks every second which we can't expect.

The results show that C5 decision tree has the most significant detection rate for both phase and level approaches. CRT & CHAID have promising results in Data Partitioning technique for both phase and level approaches.
Quest has high classification rate when adding new attacks in the second level.

The experimental results showed that Phase Model approach has Higher Classification Rate in New Attacks and Data Partitioning Techniques than Level Model approach. Therefore, the phase approach is more realistic than Level approach as in real life we are exposed every second to new attacks that we don't expect.

The Future work will be directed towards finding ways to prevent propagating errors in phase model. Also using other Machine learning techniques in our experiments for detecting more types of intrusions.

AUTHORS PROFILE

**Heba Ezzat Ibrahim** Bachelor of Computer Science. Currently working for master degree in Arab Academy for Science and Technology & Maritime Transport.

**Sherif M. Badr** PHD degree in Computer Engineering in Military Technical College. Fields of interest are intrusion detection, computer and networks security

**Mohamed A. Shaheen** Associate Professor in College of Computing and Information Technology in Arab Academy for Science and Technology & Maritime Transport